\documentclass[conference]{IEEEtran}
\usepackage{hyperref}

\ifCLASSINFOpdf
   \usepackage[pdftex]{graphicx}
 
\else
 
\fi
\pdfoutput=1
\begin{document}
%
\title{Applications of UWB Technology}

\author{\IEEEauthorblockN{Sana Ullah$^{\Psi}$ , Murad Ali$^{+}$, Md. Asdaque Hussain$^{\Psi}$, and Kyung Sup Kwak$^{\Psi}$}
\IEEEauthorblockA{$^{\Psi}$ Graduate School of IT and Telecommunications, Inha University\\
253 Yonghyun-Dong, Nam-Gu, Incheon 402-751, South Korea.\\
$^{+}$ KDI School of Public Policy and Management \\
87 Hoegiro Dondaemun, 130-868, Seoul, South Korea\\
Email: sanajcs@hotmail.com, muradjee81@hotmail.com, its\textunderscore asdaque@hotmail.com, kskwak@inha.ac.kr}}

%


\maketitle

\begin{abstract}
Recent advances in wideband impulse technology, low power communication along with unlicensed band have enabled ultra wide band (UWB) as a leading technology for future wireless applications. This paper outlines the applications of emerging UWB technology in a private and commercial sector. We further talk about UWB technology for a wireless body area network (WBAN).
\end{abstract}


%
\IEEEpeerreviewmaketitle

\section{Introduction}
There have been tremendous research efforts to apply ultra wide band (UWB) technology to the military and government sectors. Some of them are already accomplished and some are intended for future. These applications are mainly categorized into three parts: communications and sensors, position location and tracking, and radar. \\

This paper presents a brief discussion on the aforementioned applications. It is divided into five sections. Section 2 outlines the application of UWB technology in communication and sensors. Section 3 presents discussion on position location and tracking. In section 4, we talk about radar. Section 5 presents the application of UWB in a wireless body area network (WBAN). The final section presents conclusion.   

\section{Communications and Sensors}

\subsection{Low Data Rate}
The power spectral density (PSD) of UWB signals is extremely low, which enables UWB system to operate in the same spectrum with narrowband technology without causing undue interference. The solution on the market for today's indoor application is infrared or ultrasonic approaches. The line-of-sight propagation in infrared technology cannot be guaranteed all the time. It is also affected by shadows and light-related interferences. Ultra sonic approach propagates with confined penetration. UWB technology is less affected by shadows and allows the transmission through objects. The innovative communication method of UWB at low data rate gives numerous benefits to government and private sectors. For instance, the wireless connection of computer peripherals such as mouse, monitor, keyboard, joystick and printer can utilize UWB technology. UWB allows the operation of multiples devices without interference at the same time in the same space. It can be used as a communication link in a sensor network. It can also create a security bubble around a specific area to ensure security. It is the best candidate to support a variety of WBAN applications. A network of UWB sensors such as electrocardiogram (ECG), oxygen saturation sensor (SpO2) and electromyography (EMG) can be used to develop a proactive and a smart healthcare system. This can benefit the patient in chronic condition and provides long term health monitoring. In UWB system, the transmitter is often kept simpler and most of the complexity is shifted towards receiver, which permits extremely low energy consumption and thus extends battery life. 

Designing RAKE-receiver for low power devices is a complicated issue. Energy detection receivers are the best approach to build simple receivers \cite{1}. A RAKE receiver for on-body network is presented in \cite{2}, which shows that 1 or 2 fingers is sufficient to collect 50\% to 80\% of maximum energy for links with a distance of 15cm, independent of its placement on the body. Different energy management schemes may also assist in extending the battery life . Positioning with previously unattained precision, tracking, and distance measuring techniques, as well as accommodating high node densities due to the large operating bandwidth are also possible \cite{3}. Global positioning system (GPS) system is often available in low date rate applications and requires new solutions. The reduction in protocol overhead can decrease the energy consumption of the complex GPS transceivers and extends the battery life. Even though, a low data rate application using alternative PHY concepts is currently discussed in IEEE 802.15.4a \cite{4} but tremendous research efforts are required to bring those systems to real world applications. 

\subsection{High Data Rate}
The unique applications of UWB systems in different scenarios have initially drawn much attention, since many applications of UWB spans around existing market needs for high data rata applications. Demand for high density multimedia applications is increasing, which needs innovative methods to better utilize the available bandwidth. UWB system has the property to fill the available bandwidth as demand increases. The problem of designing receiver and robustness against jamming are main challenges for high-rate applications \cite{5}. The large high-resolution video screens can benefit from UWB. These devices stream video content wirelessly from video source to a wall-mounted screen. Various high data rata applications include internet access and multimedia services, wireless peripheral interfaces and location based services. Regardless of the environment, very high data rate applications ($>$1 Gbit/sec) have to be provided. The use of very large bandwidth at lower spectral efficiency has designated UWB system as a suitable candidate for high internet access and multimedia applications. The conventional narrowband system with high spectral efficiency may not be suitable for low cost and low power devices such as PDA or other handheld devices. Standardized wireless interconnection is highly desirable to replace cables and propriety plugs \cite{6}.The interconnectivity of various numbers of devices such as laptops and mobile phones is increasingly important for battery-powered devices. 

\subsection{Home Network Applications}
Home network application is a crucial factor to make pervasive home network environment. The wireless connectivity of different home electronic systems removes wiring clutter in living room. This is particularly important when we consider the bit rate needed for high definition television that is in excess of 30 Mbps over a distance of at least few meters \cite{6}. In IEEE 1394, an attempt has been made to integrate entertainment, consumer electronics and computing within a home environment. It provides isochronous mode where data delivery is guaranteed with constant transmission speed. It is important for real time applications such as video broadcasts. The required data rates and services for different devices are given in Table \ref{tab:1}. IEEE 1394 also provides asynchronous mode where data delivery is guaranteed but no guarantee is made about the time of arrival of the data \cite{7}. The isochronous data can be transferred using UWB technology. A new method which allows IEEE 1394 equipment to transfer an isochronous data using a UWB wireless communication network is presented in \cite{8}. A connection management protocol (CMP) and IEEE 1394 over UWB bridge module can exchange isochronous data through IEEE 1394 over UWB network.

\begin{table}[!t]
\renewcommand{\arraystretch}{1.3}
\caption{Contents and Requirements for Home Networking and Computing \cite{9}}
\label{tab:1}
\centering
\begin{tabular}{|c|c|c|}
\hline
Service & Data Rate (Mbps) & Real Time Feature\\
\hline
Digital Video & 32 & Yes\\
\hline
DVD, TV & 2-16 & Yes\\
\hline
Audio & 1.5 & Yes\\
\hline
PC & 32 & No\\
\hline
Internet & $>$10 & No\\
\hline
Other & $<$1 & No\\
\hline
\end{tabular}
\end{table}

\section{Position Location and Tracking}
Position location and tracking have wide range of benefits such as locating patient in case of critical condition, hikers injured in remote area, tracking cars, and managing a variety of goods in a big shopping mall. For active RF tracking and positioning applications, the short-pulse UWB techniques offer distinct advantages in precision time-of-flight measurement, multipath immunity for leading edge detection, and low prime power requirements for extended-operation RF identification (RFID) tags \cite{10}. The reason of supporting human-space intervention is to identify the persons and the objects the user aims at, and identifying the target task of the user. Knowing where a person is, we can figure out near to what or who this person is and finally make a hypothesis what the user is aiming at \cite{11}. This human-space intervention could improve quality of life when used in a WBAN. In a WBAN, a number of intelligent sensors are used to gather patient's data and forwards it to a PDA which is further forwarded to a remote server. In case of critical condition such as arrhythmic disturbances, the correct identification of patient's location could assist medical experts in treatment. 

\section{Radar}

A short-pulse UWB techniques have several radar applications such as higher range measurement accuracy and range resolution, enhanced target recognition, increased immunity to co-located radar transmissions, increased detection probability for certain classes of targets and ability to detect very slowly moving or stationary targets \cite{10}. UWB is a leading technology candidate for micro air vehicles (MAV) applications \cite{12}. The nature of creating millions of ultra-wideband pulses per second has the capability of high penetration in a wide range of materials such as building materials, concrete block, plastic and wood.
\section{Ultra Wideband Technology in WBAN}
A WBAN consists of miniaturised, low power, and non-invasive/invasive wireless biosensors, which are seamlessly placed on or implanted in human body in order to provide a smart and adaptable healthcare system. Each tiny biosensor is capable of processing its own task and communicates with a network coordinator or a PDA. The network coordinator sends patient's information to a remote server for diagnosis and prescription. A WBAN requires the resolution of many technical issues and challenges such as interoperability, QoS, scalability, design of low power RF data paths, privacy and security, low power communication protocol, information infrastructure and data integrity of the patient's medical records. The average power consumption of a radio interface in a WBAN must be reduced below $100 \mu W$. Moreover, a WBAN is a one-hop star topology where power budget of the miniaturised sensor nodes is limited while network coordinator has enough power budget. In addition, most of the complexity is shifted to the network coordinator due to its capability of having abundant power budget. The emerging UWB technology promises to satisfy the average power consumption requirement of the radio interface ($100 \mu W$), which cannot be achieved by using narrowband radio communication, and increases the operating period of sensors. In the UWB system, considerable complexity on the receiver side enables the development of ultra-low-power and low-complex UWB transmitters for uplink communication, thereby making UWB a perfect candidate for a WBAN. The difficulty in detecting noise-like behavior and robustness of UWB signals offer high security and reliability for medical applications \cite{5}.  

Existing technological growth has facilitated research in promoting UWB technology for a WBAN. The influence of human body on UWB channel is investigated in \cite{13} and results about the path loss and delay spread have been reported. The behavior of UWB antenna in a WBAN has also been presented in \cite{14}. Moreover, a UWB antenna for a WBAN operating in close vicinity to a biological tissue is proposed in \cite{15}. This antenna can be used for WBAN applications between 3 GHz and 6 GHz bands.

\begin{figure}[!t]
\centering
\includegraphics[width=3in]{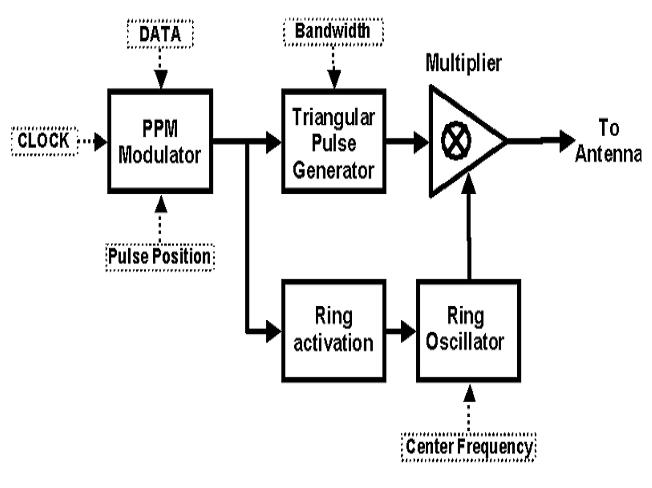}
\caption{Block Diagram of Pulse Generator \cite{16}}
\label{fig:1}
\end{figure}

A low complex UWB transmitter being presented in \cite{16} adapts a pulse-based UWB scheme where a strong duty cycle is produced by restricting the operation of the transmitter to pulse transmission. This pulse-based UWB scheme allows the system to operate in burst mode with minimal duty cycle and thus, reducing the baseline power consumption. A low power UWB transmitter for a WBAN requires the calibration of PSD inside the federal communication commission (FCC) mask for indoor application. The calibration process is a challenging task due to the discrepancies between higher and lower frequencies. Two calibration circuits are used to calibrate the spectrum inside FCC mask and to calibrate the bandwidth. A pulse generator presented in Fig. \ref{fig:1} activates triangular pulse generator and a ring oscillator simultaneously. During pulse transmission, the ring oscillator is activated by a gating circuit, thus avoiding extra power consumption. A triangular pulse is obtained at output when the triangular signal is multiplied with the output carrier produced by the oscillator. 
%
%

Finite-difference time-domain (FDTD) is used to model UWB prorogation around the human body where the path loss depends on the distance and increases with a large fading variance. The narrowband implementation is compared in terms of power consumption using a WBAN channel model. Simulation results showed that the path loss near the body was higher than the path loss in free space. Moreover, it was concluded that the performance of a UWB transmitter for a WBAN is better for best channel while for average channels narrow band implementation is a good solution \cite{16}.A considerable research efforts are required both at algorithmic and circuit level to make UWB a key technology for WBAN applications.

\section{Conclusions} 
The UWB is a leading technology for wireless applications including numerous WBAN applications. In this paper, we discussed the current and future applications of emerging UWB technology in a private and commercial sector. We believe that the UWB technology can easily satisfy the energy consumption requirements of a WBAN. Our future work includes the investigation of UWB technology for a non invasive WBAN.
\end{document}